\documentclass[aps,preprint]{revtex4}
\usepackage{graphicx}
\usepackage{bm}

\def\la{\langle}
\def\ra{\rangle}

\newcommand{\be}[1]{\begin{equation}\label{#1}}
\newcommand{\ee}{\end{equation}}
\newcommand{\ba}[1]{\begin{eqnarray}\label{#1}}
\newcommand{\ea}{\end{eqnarray}}

\begin{document}
\date{\today}
\title{The brachistochrone problem in open quantum systems 
}
\author{Ingrid Rotter}
\affiliation{Max-Planck-Institut f\"ur Physik komplexer
Systeme, D-01187 Dresden, Germany }


\begin{abstract}
Recently, the quantum brachistochrone problem is discussed 
in the literature by using  non-Hermitian Hamilton operators
of different type. 
Here, it is demonstrated that the passage time is tunable 
in realistic open quantum systems due to the biorthogonality of the
eigenfunctions of the non-Hermitian Hamilton operator. As an example,
the numerical  results obtained by Bulgakov et al. for the
transmission through  microwave cavities of different shape 
are analyzed from the point of view of the brachistochrone
problem. The passage time is shortened in the 
crossover from the weak-coupling to the strong-coupling regime
where the resonance states overlap and many branch points 
(exceptional points) in the complex plane exist. 
The effect can {\it not} be described in the framework of  standard 
quantum mechanics with  Hermitian Hamilton operator
and consideration of  $S$ matrix poles. 

\end{abstract}

\pacs{03.65.Xp, 03.65.Yz, 03.65.Vf, 73.63.Kv}
 
\maketitle

\section{Introduction}
Recently, the quantum brachistochrone problem is discussed in the literature
with great interest. It consists in finding the minimal time  
for the transition from a given initial state  $|\psi_i\rangle$
to a given  final state $|\psi_f\rangle$ with 
$|\psi_f\rangle = e^{-i\tau H}|\psi_i\rangle $.
Bender et al. \cite{bender}  found that this minimal (passage) time
can be made arbitrarily small by  parametrical variation of $H$ when   
$H$ is a non-Hermitian but ${\cal P T}$-symmetric Hamiltonian.
Assis and Fring \cite{fring} demonstrated that such a phenomenon can also 
be obtained for
dissipative systems and concluded that the effect of a tunable passage time
can be attributed to the non-Hermitian nature of the time-evolution operator
rather than to its ${\cal PT}$-symmetry. In another paper devoted 
to this topic,
Mostafazadeh \cite{ali} showed that it is  impossible to achieve faster
unitary evolutions using ${\cal PT}$-symmetric or other non-Hermitian
Hamiltonians than those given by Hermitian Hamiltonians. 
G\"unther et al. \cite{gurosa}  found that 
the passage time is reduced under the influence  of exceptional points being
branch points in the complex energy plane.  
The quantum brachistochrone problem is considered recently also by other 
authors \cite{giri,nesterov,fryd,gusa,ali2}.

The question remains open whether or not this effect is observable in a
realistic quantum mechanical system.
The best way to find an answer to this question is the consideration of 
the transmission through a quantum mechanical device such as, e.g., a 
microwave cavity. The waves propagate in the leads attached to the cavity 
which, on its part, represents an "impurity" for the propagation.  
According to standard quantum mechanics, the propagation 
of the waves  through the cavity occurs at the 
positions in energy of resonance states. 
The transmission peaks have a structure characteristic of  
resonances. The  time which the wave spends in the system,  is determined 
by the lifetime of the resonance states. That means, the transmission 
occurs via so-called {\it standing modes}.
This resonance picture of the transmission process
describes well the experimentally observed situation as long as the 
individual (long-lived)
resonance states are well isolated from one another. It breaks down,
however, in the regime of strongly overlapping resonances as numerical
studies on Sinai billiards of different shape as well as on  
quantum billiards of Bunimovich type in the framework of the tight-binding
lattice model \cite{datta} have  shown \cite{brscorr}. In this regime,
the transmission picture does not show any resonance structure.
Instead, the transmission is plateau-like as a function of energy
\cite{rs1,brsphas,brscorr}. 
It is enhanced   and the delay time (determined by the
lifetime of the resonance states) is shortened 
\cite{delay,rs1,brsphas,brscorr}.
Finally, the system becomes transparent and 
{\it traveling modes} appear inside the system.
This behavior of the transmission probability
is shown to be correlated with a reduction
of  the phase rigidity  of the scattering wave function inside the system 
and with spectroscopic reordering processes taking place in it
\cite{brsphas,brscorr}.

It is the aim of 
the present paper to show that the enhanced transmission through 
a quantum billiard in the regime of strongly overlapping resonances 
as well as the shortened
delay time accompanying it, can be traced back to
the existence of branch points \cite{excep} in the complex plane 
at which the eigenvalues of at least two eigenstates  coalesce.
Under the influence of these  points, the phases of the
eigenfunctions of the non-Hermitian  Hamilton operator 
describing the open quantum system, cease to be rigid. This behavior 
contrasts with the rigidity of the phases of the eigenfunctions of a Hermitian
Hamilton operator.  Thus,
the shortening of the evolution time in physical systems 
whose states are described by a non-Hermitian Hamilton operator
is a realistic effect, indeed,  and can be observed in
realistic open quantum systems in the regime of overlapping resonances.

In Sect. 2 of the present paper, the appearance of the non-Hermitian Hamilton
operator $H_{\rm eff}$ in the Feshbach projection operator (FPO) technique
is sketched. In Sect. 3, the mathematical freedom in the  normalization 
of the eigenfunctions $\phi_\lambda$ of $H_{\rm eff}$ is discussed as well as 
the consequences of the chosen normalization for the rigidity 
of the phases of the $\phi_\lambda$. The phase rigidity $r_\lambda$
is introduced and compared with the results of an experimental study
performed on a microwave cavity. In the next section (Sect. 4), 
the phase rigidity 
$\rho$ of the scattering wave function inside the system is defined.
At the considered energy $E$ of the system
it contains, in the regime of overlapping resonances, 
contributions of all the individual resonance states
$\lambda$ including their phase rigidities $r_\lambda$. 
The results of the $S$ matrix theory for the
transmission through a microwave cavity are sketched in Sect. 5.
Deviations from the standard theory based on 
the resonance structure of the transmission, appear only in the regime of
overlapping resonances. Here, the phase rigidity $\rho$ is reduced 
due to the many branch points characteristic of this regime.
It may happen $\rho \to 0$ in a broader parameter range. In such a case, the 
transmission has a  plateau-like structure and
may occur  via traveling modes. The system becomes transparent.
In Sect. 6, the results are summarized.and some conclusions are drawn.

\section{Feshbach projection operator (FPO) technique}

In the present paper, the FPO technique \cite{feshbach} will be used 
in order to describe the transmission through an open quantum microwave cavity.
In the FPO formalism, the full function space is divided into two subspaces:
the $Q$ subspace contains all wave functions that are localized inside the 
system and vanish outside of it while
the wave functions of the $P$ subspace are extended up to infinity and vanish
inside the system, see \cite{rep}. It is $P+Q=1$.
In this formalism, two Hamilton operators characterize the system. 
The first one, $H$, is Hermitian. It describes the
scattering in the whole function space, 
\begin{eqnarray}
(H-E)\;\Psi^E_C = 0, 
\label{Psi}
\end{eqnarray}
consisting of the two subspaces: the subspace of discrete states of the 
considered (closed) system (described by the Hermitian operator $H_B$)
and of the subspace of scattering states (continuum described by 
the Hermitian operator $H_C$) into which the system is embedded. 
In solving (\ref{Psi}) in the whole function space by using the 
FPO technique \cite{feshbach}, the effective non-Hermitian Hamilton operator 
\begin{eqnarray}
H_{\rm eff} = H_B + \sum_C V_{BC} \frac{1}{E^+ - H_C} V_{CB} 
\label{heff}
\end{eqnarray}  
appears which contains $H_B$ as well as an additional
symmetrical non-Hermitian term 
that describes the coupling of the resonance states via the common environment.
Here  $V_{BC}, ~V_{CB}$ stand for the coupling matrix elements between the 
{\it eigenstates} of $H_B$ and the environment \cite{rep}
that may consist of different continua $C$. The operator $H_{\rm eff}$
is non-Hermitian,
\begin{eqnarray}
(H_{\rm eff} - z_\lambda)\,\phi_\lambda =0 \; ,
\label{phi}
\end{eqnarray}
its eigenvalues $z_\lambda$ and eigenfunctions 
$\phi_\lambda$ are complex. The eigenvalues
provide not only the energies of the resonance
states but also their widths (inverse lifetimes). The eigenfunctions
are biorthogonal.

The eigenvalues and eigenfunctions of $H_B$ contain the  
interaction $u$ of the discrete states which is given by the 
nondiagonal matrix elements of $H_B$. This interaction 
is of standard type in closed systems and may be called therefore
internal interaction. The
eigenvalues and eigenfunctions of $H_{\rm eff}$ contain additionally the
interaction $v$ of the resonance states via the 
common continuum ($v$ is used here instead of the concrete 
matrix elements of the second term of $H_{\rm eff}$). 
This part of interaction is, formally, of second order and
may be called  external interaction.
While $u$ and Re$(v)$ cause  level repulsion in energy, 
Im$(v)$ is responsible for the bifurcation of the widths 
of the resonance states  (resonance trapping). 
The phenomenon of resonance trapping appearing in the regime of overlapping
resonances, has been proven experimentally in a microwave cavity
\cite{stm}.

Since the effective Hamilton operator (\ref{heff}) depends explicitly  on 
the energy $E$, so do its eigenvalues $z_\lambda$ and eigenfunctions
$\phi_\lambda$. Far from thresholds, 
the energy dependence 
is weak, as a rule, in an energy interval of the order of magnitude of the 
width of the resonance state. 
The solutions of the fixed-point equations
$E_\lambda={\rm Re}(z_\lambda)_{|E=E_\lambda}  $ and of
$\Gamma_\lambda=-2\, {\rm Im}(z_\lambda)_{|E=E_\lambda} $
are numbers that coincide  with the poles of the $S$ matrix. 
The widths $\Gamma_\lambda$ determine
the time scale characteristic of the resonance states $\lambda$.
Using the FPO formalism with non-Hermitian Hamilton operator $H_{\rm eff}$,
it is however not necessary to look for the poles of the $S$ matrix 
since the  spectroscopic information is involved in  the
complex eigenvalues $z_\lambda$ of $H_{\rm eff}$.
In the $S$ matrix,  the eigenvalues $z_\lambda$ with their full 
energy dependence appear. Due to this fact, the 
$S$ matrix  contains information on the environment of the 
considered resonance states such as the position of decay
thresholds and of neighboring resonance states.

Thus, the FPO formalism may be considered as an extension \cite{acta}
of the  $R$ matrix theory used in standard quantum mechanics
for the description of decaying states.
The standard spectroscopic parameters (positions, 
widths and partial widths of the resonance states $\lambda$)  
are replaced by the energy dependent 
functions $E_\lambda, ~\Gamma_\lambda$ and coupling matrix elements between 
system and environment. While $R$ matrix theory gives reasonable 
results only for narrow non-overlapping resonance states, 
the FPO formalism can be used for all resonance states including the broad 
ones in the overlapping regime. The influence of neighboring resonances as
well as of decay thresholds is taken into account via the energy dependence of
the eigenvalues $z_\lambda$ and eigenfunctions $\phi_\lambda$.
The spectroscopic information can be controlled by means of an external 
parameter. Also the redistribution processes taking place under the influence
of the coupling to the continuum in the overlapping regime can be traced. 
The  results obtained in the  FPO 
formalism pass into those of the $R$ matrix theory when the 
overlapping of the resonance states vanishes.

\section{Phase rigidity of the eigenfunctions
$\phi_\lambda$ of the non-Hermitian Hamilton operator $H_{\rm eff}$}

The eigenfunctions $\phi_\lambda$  of the non-Hermitian  
symmetrical Hamilton operator $H_{\rm eff}$
are complex and  biorthogonal.
The normalization condition $\langle \phi_\lambda^{\rm left} |
\phi_\lambda^{\rm right}\rangle =     
\langle\phi_\lambda^*|\phi_{\lambda }
\rangle $ fixes only two of the four free parameters \cite{gurosa}. 
This freedom can be used  in order to provide a smooth transition from 
an open quantum system (with, in general, nonvanishing decay widths
$\Gamma_\lambda$ of its states and biorthogonal wave functions
$\phi_\lambda $) to the corresponding closed one (with 
$\Gamma_\lambda \to 0$ and real wave functions that are normalized
in the standard manner): $\langle\phi_\lambda^*|\phi_{\lambda }
\rangle \to \langle\phi_\lambda|\phi_{\lambda }
\rangle =1 $ if the coupling vectors in the non-Hermitian part of 
(\ref{heff}) vanish.
That means, the orthonormality conditions can be chosen as
\begin{eqnarray} 
\langle\phi_\lambda^*|\phi_{\lambda '}
\rangle = \delta_{\lambda, \lambda '} 
\label{biorth1} 
\end{eqnarray}
with the consequence that \cite{rep}
\begin{eqnarray}
\langle\phi_\lambda|\phi_{\lambda}\rangle  \equiv  A_\lambda \ge 1
\label{biorth2a} \\
B_\lambda^{\lambda '} \equiv 
\langle\phi_\lambda|\phi_{\lambda ' \ne \lambda}\rangle = -B_{\lambda
'}^\lambda \equiv  
- \,\langle\phi_{\lambda '\ne \lambda }|\phi_{\lambda }\rangle 
\nonumber \\
\qquad  \qquad |B_\lambda^{\lambda '}|
~\ge ~0  \; .
\label{biorth2b}
\end{eqnarray}
The normalization condition (\ref{biorth1}) entails that 
the phases of the eigenfunctions in the overlapping regime
are not rigid: the normalization condition
$\la\phi_\lambda^*|\phi_{\lambda}\ra =1$ is fulfilled 
only when Im$\langle \phi_\lambda^*|\phi_\lambda\rangle \propto $
Re$~\phi_\lambda \cdot$  Im$~\phi_\lambda =0$, i.e. 
by rotating the wave function at a certain angle $\beta_\lambda$. 
The phases of the wave functions of the original states 
corresponding to $v=0$ [vanishing non-diagonal matrix elements 
of the second term  of (\ref{heff})] are fixed, say 
to $\beta_\lambda^{0} = 0$ or $\pm\pi$, so that Im$\,\phi_\lambda^0 =0$. The
influence of a neighboring state is described by  $v\ne 0$ [i.e. 
by the  non-diagonal matrix elements of 
the second term of (\ref{heff})]. 
At $v\ne 0$,  the angle  $\beta_\lambda$  is different from
$\beta_\lambda^{0}$, generally. The difference $|\beta_\lambda -
\beta_\lambda^0|$  may be $\pm \pi /4$ at most, corresponding to
Re$~\phi_\lambda = \pm \, {\rm Im}~\phi_\lambda$
(as compared to Im$\,\phi_R^0 =0$). This maximum value
appears at a branch point in the complex energy plane 
\cite{excep} where 
two eigenvalues  of $H_{\rm eff}$ coalesce \cite{gurosa,rep,rs1}. 
Here \cite{rep,rs2,gurosa}
\begin{eqnarray}
\phi_\lambda \to \pm\, i\, \phi_{\lambda '} \; \; ;
\qquad  \phi_{\lambda '} \to \mp \, i \, \phi_\lambda \; .
\label{r1a} 
\end{eqnarray}
The phase rigidity defined by
\begin{eqnarray}
r_\lambda = 
\frac{\langle \phi_\lambda^*| \phi_\lambda \rangle }
{\langle\phi_\lambda | \phi_\lambda \rangle} = 
\frac{1}{({\rm Re}\, \phi_\lambda) ^2 + ({\rm Im}\, \phi_\lambda) ^2}= 
\frac{1}{A_\lambda}
\label{ph2} 
\end{eqnarray}
is a useful measure \cite{brsphas} for the 
rotation angle $\beta_\lambda$. 
When the resonance states are distant from one another, it is 
$r_\lambda \approx 1$ due to $ \langle\phi_\lambda |\phi_\lambda\rangle$ 
$\approx \langle\phi_\lambda^*|\phi_\lambda\rangle$. 
In approaching a branch point in the complex energy plane \cite{rs2,rep}, 
we have  
$~\langle\phi_\lambda | \phi_\lambda \rangle \equiv A_\lambda \to \infty$ 
and  $r_\lambda \to 0$. Therefore $1\ge r_\lambda \ge 0$. 

It should be underlined  that,
after defining the normalization condition (\ref{biorth1}), 
the values $r_\lambda$ are fixed by the coupling matrix
elements $v$ of $H_{\rm eff}$ characteristic of the degree of overlapping 
of the resonance states. They can be varied by controlling the system by means
of external parameters, e.g. by means of a laser in the case of an atom
with many levels (for concrete examples  see \cite{marost23}).
The rotation angle $\beta_\lambda$ as well as the
values $A_\lambda$ and $r_\lambda$ may be considered to be a
synonym for the biorthogonality of the eigenfunctions $\phi_\lambda$
of the non-Hermitian Hamiltonian (\ref{heff}). They are a measure for the
distance of the considered states from a branch point in the complex
plane and for the spectroscopic reordering processes occurring in an open
quantum system with overlapping resonance states
under the influence of the coupling to the continuum.
Physically, the phase
rigidity $r_\lambda$ measures the degree of alignment of 
one of the neighboring resonance states with one of the scattering states 
$\xi^E_C$ of the environment. This alignment takes place at the cost of the
other states that decouple, to a certain extent,  from the 
environment ({\it widths bifurcation} or {\it resonance trapping} occurring in
the neighborhood  of a branch point in the complex energy plane \cite{rep}). 
The $r_\lambda$ are, generally, different for the different states $\lambda$.

We consider now
the experimental results obtained on a microwave cavity \cite{demb2}. 
The experimental conditions are chosen in such a manner that the
phase difference between the oscillating fields at the position of the 
antennas is $\pi$ far from the branch point \cite{excep}. 
Then the phase difference 
is traced experimentally in approaching the branch point: in a 
comparably large parameter range,
it  drops eventually to $\pi/2$ at the branch point. For an interpretation
of the results, the  authors  \cite{demb2} consider the reduced
phase difference only at the branch point and relate it to the existence of 
a chiral state. They do not discuss the smooth reduction 
from $\pi$ to almost $\pi /2$ in approaching the branch point. 

According to the  discussion above, the 
experimentally observed \cite{demb2} reduction of the
phase difference between the wave functions of the two states can be 
related to the reduction of the phase rigidity of the two wave functions.
The phase rigidity drops smoothly from its maximum value
$r_\pm =1$ far from the branch point 
[with the phase difference $\pi$ (or $2\pi$) 
between the  wave functions of isolated resonance states]   
to its minimum value $r_\pm =0$ at the branch point
[with the phase difference $\pm\pi /2$ according to (\ref{r1a})].
This interpretation  explains, in a natural manner, 
the experimentally observed smooth reduction of the phase difference 
in a comparably large parameter range. Also the phase jump occurring at the
branch point \cite{gurosa} is not in disagreement with the experimental data. 
The results demonstrate the (parametric) dynamics of  open quantum systems
which is generated by the interaction of resonance states via the continuum
as discussed above.

\section{Phase rigidity of the scattering wave function $\Psi^E_C$}

The solution of the whole problem  (\ref{Psi})
with the Hermitian Hamilton operator $H$ reads \cite{rep}
\begin{equation}
\Psi_C^E= \xi^E_C + \sum_{\lambda}
\Omega_\lambda^C
~\frac{\langle\phi_\lambda^* |V| \xi^E_C\rangle}{E-z_\lambda} 
\label{total}
\end{equation}
where
\begin{equation}
\label{reswf}
\Omega_\lambda^C= 
\Big( 1+ \frac{1}{E^{+}-H_C} V_{CB}\Big)\phi_\lambda 
\end{equation}
is the wave function of the resonance state $\lambda$ and
the $\xi^E_C$ are the (coupled) scattering wave functions 
of the continuum into which the system is embedded.
According to (\ref{total}), the eigenfunctions $\phi_\lambda$ of the 
non-Hermitian Hamilton operator $H_{\rm eff}$ give the main
contribution to the scattering wave function $\hat\Psi^E_C$ in the interior 
of the system,
\begin{eqnarray}
\Psi_C^E \to \hat \Psi_C^E =
\sum_\lambda c_{\lambda E}\, \phi_\lambda  \, ; 
\quad \; c_{\lambda E} =
\frac{\langle \phi_\lambda^* |V| \xi^E_C\rangle}{E-z_\lambda } \; .
\label{total1}
\end{eqnarray}
The weight factors $c_{\lambda E}$
contain the excitation probability of the states $\lambda$.

In the FPO method supplemented by the normalization condition (\ref{biorth1}),
the definition of the two subspaces (system and
environment) appears in a natural manner: $H_B$ describes the closed system
which becomes open when embedded in the continuum of scattering wave
functions $\xi^E_C$ described by $H_C$. Therefore, all spectroscopic 
values characteristic of
resonance states can be traced to the corresponding values of discrete states
by controlling the coupling to the continuum. That means, with $v\to 0$, 
the transition from resonance states (described by the non-Hermitian 
$H_{\rm eff}$) to discrete states (described by the Hermitian $H_B$)
can be controlled.  

Let us consider the one-channel case, $C=1$, and $\Psi^E_C \to 
\hat\Psi^E$
in the interior of the system. From (\ref{total1})
follows for the right and left wave functions
\begin{eqnarray}
|\hat\Psi^{E, R}\ra=  \sum_\lambda c_{\lambda E}\, |\phi_\lambda^R\ra  
\label{r1} \\
\la\hat\Psi^{E, L} |=  \sum_\lambda d_{\lambda E} \, \la\phi_\lambda^L |
\label{r2}
\end{eqnarray}
with $|\phi_\lambda^R\ra \equiv |\phi_\lambda\ra$, 
$\la\phi_\lambda^L | =  \la \phi_\lambda^* | $ and $d_{\lambda E} = 
c_{\lambda E}^*$ when excitation and decay of the state $\lambda$ 
occur via the same mechanism. Therefore the $\hat\Psi^{E}$ 
can be normalized,
\begin{eqnarray}
\la\hat\Psi^{E, L} | \hat\Psi^{E, R}\ra & = 
& \sum_{\lambda \lambda '}
c_{\lambda E}^* c_{\lambda ' E} \; \la\phi_\lambda^* |\phi_{\lambda '}\ra
\nonumber \\  
&=& \sum_\lambda |c_{\lambda E}|^2 \equiv 1 \; .
\label{r4}
\end{eqnarray}
The normalization has to be done separately
at every energy $E$ due to the 
explicit energy dependence of the  $c_{\lambda E}$. 
Moreover,
\begin{eqnarray}
\la\hat\Psi^{E, L\,*} | \hat\Psi^{E, R}\ra
&=& \sum_{\lambda \lambda '}
c_{\lambda E} c_{\lambda ' E} \; \la\phi_\lambda |\phi_{\lambda '}\ra
\nonumber \\  
&=& \sum_\lambda (c_{\lambda E})^2 \, A_\lambda + \sum_{\lambda < \lambda '}
c_{\lambda E} c_{\lambda ' E} \, (B_\lambda^{\lambda '} + 
B_{\lambda '}^\lambda)
\nonumber \\
&=& \sum_\lambda (c_{\lambda E})^2 \, A_\lambda 
\label{r5}
\end{eqnarray}
due to $B_\lambda^{\lambda '} = - B_{\lambda '}^\lambda$, see (\ref{biorth2b}).
$A_\lambda$ is a real number, see \cite{rep}.
From (\ref{r4}) and (\ref{r5}) follows
\begin{eqnarray}
\frac{\langle \hat\Psi^{E*} | 
\hat\Psi^{E} \rangle}{\la\hat\Psi^{E} | \hat\Psi^{E}\ra}
&=& \sum_\lambda (c_{\lambda E})^2\, A_\lambda
= \sum_\lambda \frac{(c_{\lambda E})^2}{ r_\lambda} \; , 
\label{r6}
\end{eqnarray}
and the phase rigidity $\rho$
of the wavefunctions $\hat\Psi^E$ may be defined by 
\begin{eqnarray}
\rho &=& 
e^{2i\theta} \sum_\lambda\frac{{\rm Re}\,[(c_{\lambda E})^2]}{r_\lambda} 
\nonumber \\
&=&
e^{2i\theta} \sum_\lambda \frac{1}{r_\lambda}
\bigg( [{\rm Re}(c_{\lambda E})]^2 - [{\rm Im}(c_{\lambda E})]^2 \bigg) 
\label{r7}
\end{eqnarray}
in analogy to (\ref{ph2}).
The value $\rho$ corresponds to a rotation of $\hat\Psi^E$ 
by $\theta$ corresponding to the ratio between its real and imaginary
parts.  
In spite of the complicated structure of $\rho$, it holds $1 \ge \rho \ge 0$
[since  $1\le (a^2-b^2)/(a^2+b^2)\le 0 $ for every summand
$(a+ib)^2$ in (\ref{r7})].  
Eqs. (\ref{r5}) and (\ref{r7}) show that the definition of $\rho$ 
is meaningful only when the sum of all the overlapping
states $\lambda$ at the energy  $E$ of the system
is considered. The value $\rho$ is uniquely
determined by the spectroscopic properties of the system that are expressed 
by the coupling coefficients to the environment and the level density, or 
by the  positions and widths of the resonance states and the phase 
rigidities $r_\lambda$. 

According to (\ref{r7}), we have the following  border cases.
\begin{enumerate}
\item
The resonances are well separated from one another,
$\Gamma_\lambda \ll \Delta E \equiv E_\lambda -E_{\lambda '}$ : 
$r_\lambda \approx 1$ and $(c_{\lambda E})^2 \approx |c_{\lambda E}|^2 =1$
for $E\to E_\lambda$. In such a case $|\rho | \to 1$. 
\item
The resonances overlap and $r_\lambda < 1$ (but different from 0) 
for a certain number of neighboring
resonances: it may happen that $\rho = 0$ in a finite energy interval, 
see \cite{rs1,brsphas,brscorr} for numerical examples.
\item
The eigenvalues $z_\lambda$ of two resonance states coalesce at 
$E\to E_\lambda$: $r_\lambda \to 0$ and $(c_{\lambda E})^2 \to 0$ 
at this energy, see e.g. \cite{rep}. 
Therefore $\rho$ is finite at $E\to E_\lambda$. The results of a numerical 
example (double quantum dot)  are shown in \cite{brsphas}, Fig. 2.  
\item
$K$ out of $N$ wave functions $\Psi_C^E$ are aligned with the 
$K$ scattering wave functions $\xi^E_C$ of the environment while the 
remaining $N-K$ wave functions are more or less decoupled from
the continuum and well separated from one another. 
In such a case, $|\rho | \to 1$. In difference to the first case,
the $N-K$ trapped (narrow) resonance states are superposed by a background
term that arises from the $K$ aligned (short-lived) resonance states. 
\end{enumerate}
This behavior of the phase rigidity $\rho$ is traced in a numerical study 
for different quantum billiards \cite{brscorr}.
When the beam is fully reflected, it
may, of course, also happen that $|\rho | \to 1$ in a finite
energy interval. 

The wave functions $\hat\Psi^{E}$ are the exact solutions of the 
Schr\"odinger equation (\ref{Psi}) in the interior of the system. 
Eq. (\ref{r7}) shows that $\rho$ obtained for these wave functions
is related to the individual $r_\lambda$. This relation becomes important 
only in the regime of overlapping resonances where $r_\lambda < 1$.
Every value $r_\lambda$ as well as every coefficient 
$c_{\lambda E}$ are given by the concrete values of
the coupling strength between system and environment
in the concrete situation considered. 
Thus, also $\rho$ is uniquely determined by the 
degree of overlapping of the resonance states by which 
the coupling matrix elements are determined. 

This  result is in contrast to the definition of 
$\rho_{\rm br}$ given by Brouwer \cite{brouwer} by means of an arbitrary 
wave function $\tilde\Psi$ although 
\begin{eqnarray}
\rho_{\rm br}  &=&
e^{2i\Theta} \frac{\int dr (|{\rm Re}\tilde\Psi(r)|^2 - 
|{\rm Im}\tilde\Psi(r)|^2)}
{\int dr (|{\rm Re}\tilde\Psi(r)|^2 + |{\rm Im}\tilde\Psi(r)|^2)}
\label{r8}
\end{eqnarray}
is formally analog to the definition (\ref{r7}).
In the  case of $\rho_{\rm br}$, the source for the reduction 
of the phase rigidity is not known. It is rather expressed 
quite generally by the value $\rho_{\rm br}$ in analyzing 
experimental data. Unlike $\rho_{\rm br}$, the only source for the reduced
phase rigidity   (\ref{r7})  is the biorthogonality of the eigenfunctions
$\phi_\lambda$ of the non-Hermitian Hamilton operator $H_{\rm eff}$
by which the alignment of  
individual wave functions $\phi_\lambda$ with the scattering wave functions 
$\xi^E_C$ of the environment becomes possible. It can be calculated as shown,
e.g., in \cite{brsphas,brscorr}.
The alignment may be characterized by the 
corresponding rotation angles $\beta_\lambda$ or the phase rigidities
$r_\lambda$ as discussed in Sect. 3.
This effect is characteristic of open quantum systems
in the regime of overlapping resonances.  It appears also at zero temperature.

\section{Transmission through a microwave cavity}

\subsection{Isolated resonances}

According to the $S$ matrix theory, the
amplitude for the transmission through a quantum dot is \cite{sr03}
\begin{equation}
t=-2\pi
i\sum_{\lambda}\frac{\langle \xi^E_L|V|\phi_\lambda\rangle
\langle\phi_\lambda^*|V|\xi^E_R\rangle }{E-z_{\lambda}} \; . 
\label{trHeff}
\end{equation}
The eigenvalues $z_\lambda$ and eigenfunctions $\phi_\lambda$ of $H_{\rm eff}$
are involved in  (\ref{trHeff}) with their full energy dependence.

For $\rho =1$ and well isolated resonance states, 
the transmission amplitude (\ref{trHeff}) 
repeats the resonance structure of (\ref{total}) of the wave function
$\Psi^E_C$.  The transmission peaks appear at the positions 
$E_\lambda \equiv {\rm Re}(z_\lambda)_{|E=E_\lambda} \approx E_\lambda^B$ 
of the resonance states. Using the relation \cite{rep}
\begin{eqnarray}
\Gamma_\lambda & = & 2 \pi \{\langle
\xi^E_L|V|\phi_\lambda\rangle\langle\phi_\lambda^*|V|\xi^{E}_L\rangle   
+\langle
\xi^E_R|V|\phi_\lambda\rangle\langle\phi_\lambda^*|V|\xi^{E}_R\rangle \} 
\nonumber \\
&=& 4 \pi \langle
\xi^E_C|V|\phi_\lambda\rangle\langle\phi_\lambda^*|V|\xi^{E}_C\rangle
\label{sm1}
\end{eqnarray}
for the  case of a symmetrical cavity with isolated resonance states and
one channel in each of the two identical (semi-infinite) leads
($C=L,R$, respectively),  the peak height is  
\begin{eqnarray}
|t_{(E\to E_\lambda)}| &= & \frac{4\pi}{\Gamma_\lambda} \, 
|\langle \xi^E_L|V|\phi_\lambda\rangle
\langle\phi_\lambda^*|V|\xi^{E}_R\rangle| =  1 \; .
\label{sm2}
\end{eqnarray}
Except for threshold effects, the profile of the transmission peak is 
of Breit-Wigner type, determined by the width 
$\Gamma_\lambda \equiv -~2 ~{\rm Im}(z_\lambda)_{|E=E_\lambda}$ 
of the resonance state $\lambda$. 

An analogous result holds  when there is a nonvanishing background term
additional to the resonance term (\ref{trHeff}) of the transmission amplitude.
Such a term is  caused by the contribution of the scattering wave functions 
$\xi^E_C$ in (\ref{total}) to the transmission. It describes a wave
traveling through the cavity. The  time scale corresponding to this 
so-called {\it direct} part of the transmission is, generally, well separated
from that corresponding to the resonance part described by (\ref{trHeff})
\cite{comm1}.
Mostly, the resonances are  narrow and well separated from one another. 
They   appear as Fano resonances \cite{fano} on the smooth background
(caused by the traveling mode $\xi^E_C$). Due to the different time scales of
the resonance and direct processes, it is $|\rho |\approx 1$ 
also in this case.  

Thus, the resonance structure of the transmission amplitude 
with and without contributions from the direct  reaction part
can be described
in the framework of standard quantum mechanics with Hermitian Hamilton
operator and consideration of $S$ matrix poles
since the phases of the wave functions of the resonance states
are almost rigid, $\rho \approx 1$.

\subsection{Overlapping resonances}

The situation is another one when the resonances overlap.
In the overlapping regime, the resonance
states avoid crossings  with  neighbored resonance states. 
In contrast to (\ref{sm1}), it holds 
\begin{equation}
\Gamma_\lambda <  4 \pi ~\langle
\xi^E_C|V|\phi_\lambda\rangle\langle\phi_\lambda^*|V|\xi^{E}_C\rangle   
\label{sm1o}
\end{equation}
in the case with one channel in each of the two identical leads due to the 
biorthogonality of the eigenfunctions $\phi_\lambda$ \cite{rep}.
At $E\to E_\lambda$, the transmission amplitude is
\begin{eqnarray}
t_{(E\to E_\lambda)} &= & - 2 \pi i 
\sum_{\lambda ' \ne \lambda }
\frac{\langle \xi^E_L|V|\phi_{\lambda^{'}}\rangle
\langle\phi_{\lambda^{'}}^*|V|\xi^{E}_R\rangle}{E-z_{\lambda^{'}}} 
\nonumber \\ && - 4 \pi   
~\frac{\langle \xi^E_L|V|\phi_{\lambda}\rangle
\langle\phi_{\lambda}^*|V|\xi^{E}_R\rangle}{\Gamma_{\lambda}} 
 \; .
\label{sm2a}
\end{eqnarray}
It follows from (\ref{sm1o}) that the contribution of the state 
$\lambda$ to $t_{(E\to E_\lambda)}$  is larger than 1. The unitarity 
condition will be fulfilled, nevertheless, due to the
possibility to  rotate the
$\phi_\lambda$, i.e. due to phase changes of the wave functions
$\phi_\lambda$. Moreover, 
also the minima in the transmission between two 
resonance peaks may be filled up due to phase changes of the wave functions
$\phi_{\lambda }$ and $\phi_{\lambda '}$ of the two neighboring resonance 
states $\lambda$ and $\lambda '$ ~\cite{rep}. 
As a consequence,
the transmission in the overlapping regime does not show a resonance 
structure. Instead, it might be nearly plateau-like.
Let us rewrite therefore the transmission amplitude 
(\ref{trHeff})  by means of the scattering wave function (\ref{total1}),
\begin{eqnarray}
t = - 2\pi i
~\langle \xi^E_L|V|\hat\Psi^E_R\rangle 
\label{tr}
\end{eqnarray}
with $\hat\Psi^E_R$ being complex, in general.
The advantage of this representation consists in the fact that it does not
suggest the existence of resonance peaks in the transmission probability. 
Quite the contrary, the transmission is determined by the degree of alignment
of the wave function $\hat \Psi_C^E$ with the propagating modes 
$\xi^E_C$ in the leads, i.e. by the value 
$\langle \xi^E_C|V|\hat\Psi^E_C\rangle $. Nevertheless, the expressions
(\ref{tr}) and (\ref{trHeff}) are fully equivalent.

The plateau-like structure of the transmission can not be  
obtained in standard quantum mechanics with fixed phases of the 
wave functions, $r_\lambda =1$ and $\rho =1$. It is generated by 
the interference processes with account of the
alignment of some of the resonance states to the scattering 
states $\xi^E_C$  of the
environment. At most, ${\rm Re}\,\hat\Psi^E_C = \pm{\rm Im}\,\hat\Psi^E_C$
(as for the $\xi^E_C$). This case corresponds to $\rho =0$. 
It will be reached when many resonance states are almost aligned with the
$\xi^E_C$ and $\sum_\lambda {\rm Re}[(c_{\lambda E})^2] /r_\lambda
\approx 0$ according to (\ref{r7}).

Let us now 
consider  the case of two resonance states with extremely strong overlapping  
(corresponding to $r_{\lambda_1}=  r_{\lambda_2} =0$) which occurs
at the branch point in the complex energy plane. Here 
two eigenvalues $z_1$ and $z_2$ of $H_{\rm eff}$ coalesce,  $E_{\lambda_1} =  
E_{\lambda_2}\equiv E_{\lambda} $,
~$\Gamma_{\lambda_1} = \Gamma
 _{\lambda_2}\equiv \Gamma_{\lambda}$. 
In the case of one channel in each of the two identical leads,
it follows from (\ref{trHeff})
\begin{eqnarray}
t_{(E\to E_{\lambda})} & \to &  \frac{4\pi}{\Gamma_\lambda}~
\bigg(\langle \xi^E_L|V|\phi_{\lambda_1}\rangle
\langle\phi_{\lambda_1}^*|V|\xi^E_R\rangle +
~\langle \xi^E_L|V|\phi_{\lambda_2}\rangle
\langle\phi_{\lambda_2}^*|V|\xi^E_R\rangle \bigg) 
\nonumber \\ 
& & = ~0
\label{sm4}
\end{eqnarray}
at $E\to E_\lambda$ due to 
$|\phi_{\lambda}\rangle \to  \pm \; i\; |\phi_{\lambda '\ne \lambda}\rangle $
at the branch point, Eq. (\ref{r1a}).
That means, the transmission vanishes 
at the energy $E=E_\lambda$ of the two resonance states. 
The transmission profile  can be derived from (\ref{trHeff}) 
by taking into account the resonance behavior of the coupling coefficients 
of the two resonance states \cite{ro03,marost4},
\begin{eqnarray}
t= -2\, i ~\frac{\Gamma_\lambda}{E-E_\lambda+\frac{i}{2}\Gamma_\lambda}- 
\bigg(\frac{\Gamma_\lambda}{E-E_\lambda +\frac{i}{2}\Gamma_\lambda}\bigg)^2
\; .
\label{nl6}
\end{eqnarray}
The interference between both terms in (\ref{nl6}) 
causes two transmission peaks
in an energy region $\Delta E$ that is characteristic of the first term 
of (\ref{nl6}). The resulting "antiresonance" at $E=E_\lambda$  is  
narrower than  a Breit-Wigner resonance, and the two transmission peaks
are non-symmetrical. Let us compare the transmission in the energy region 
$\Delta E$ when (i) there are two coalesced eigenvalues of $H_{\rm eff}$ 
as discussed above and (ii) there are two (more or less) isolated 
resonance states resulting in two symmetrical
transmission peaks of Breit-Wigner shape. 
In both cases we have two transmission peaks, however with
a different profile. As a consequence,
the transmission is different in the two cases. It is
larger in the first case than in the second one. 
The difference is however small. 

Thus, $\rho \ne 0$ in accordance with (\ref{r7})
in spite of $r_\lambda =0 $ for the two states whose
eigenvalues coalesce at the branch point in the complex plane. 
Due to the reduced phase rigidities
$r_\lambda$ of the two states $\lambda_1$ and $\lambda_2$, this case
can be described by standard quantum mechanics with Hermitian Hamilton
operators and fixed phases of its states at the best in an approximate manner.

\subsection{Relation to the phase rigidity  $\rho$}

As a result of the above discussion, we have the following cases.
\begin{enumerate}
\item
The phases of the eigenfunctions $\phi_\lambda$ of the non-Hermitian Hamilton
operator $H_{\rm eff}$ are (almost) rigid, $|r_\lambda | \approx 1$ and 
$|\rho | \approx 1$.\\
In this case, the transmission can be described quite well by standard quantum
mechanics with a Hermitian Hamilton operator
and fixed phases. The transmission shows a
resonance structure according to the standing waves in the cavity. 
The time delay of the transmission inside the cavity
is caused by the finite lifetime of the individual resonance states.  
  
\item
The phases of the eigenfunctions  $\phi_\lambda$ of the non-Hermitian Hamilton
operator $H_{\rm eff}$ are not rigid, $|r_\lambda | <1$ and $|\rho | <1$.\\
In this case, the transmission can not be described by standard quantum
mechanics with a Hermitian Hamilton operator, 
and the transmission does not show
any pronounced resonance structure. In a comparably large parameter range,
it is rather plateau-like and the transmission occurs, in this parameter 
range,  via traveling modes through the cavity, i.e. the cavity
becomes transparent. The transmission does not occur through individual
resonance states in this case. Instead, the overlapping of the resonance
states allows the alignment of
some of them  with the traveling (scattering) states 
of the environment so that the cavity does not cause a  time delay
of the transmission. 

\end{enumerate}

The numerical results \cite{brscorr}
obtained for the transmission through microwave cavities
of different shape show exactly the features discussed above. 
In the weak-coupling
regime as well as in the strong-coupling regime, the transmission shows a
resonance structure as expected from the standard quantum mechanics with a
Hermitian Hamilton operator. The only difference between the two cases is the
appearance of a smooth background term in the strong-coupling regime
which does not exist in the weak-coupling case, and the reduction of the number
of resonance peaks by two (corresponding to the alignment of two resonance
states each with one channel in each of the
two identical attached leads).

In the crossover from the weak-coupling regime to the strong-coupling 
one, however, the transmission 
is plateau-like instead of showing a resonance structure. It 
is enhanced as compared to the transmission probability in the 
two borderline cases. In this regime, the resonance states overlap and 
spectroscopic reordering processes take place.
Due to widths bifurcation, some of the resonance states
become short-lived while other ones become trapped (long-lived).  
The enhancement of the transmission is caused by the short-lived states.
Most interesting is the
correlation between transmission $|t|$ and reduced phase rigidity $1-|\rho|$
which can be seen  very clearly in all the numerical results
shown in \cite{brscorr}. The transmission in the crossover regime
is not only enhanced but it also outspeeds the transmission
calculated in standard quantum mechanics. The reason is 
the formation of aligned (short-lived) resonance states 
in the neighborhood of branch points in the complex plane.

The behavior of the transmission in the crossover
regime with overlapping resonance states does {\it not} correspond to the
expectations of the standard quantum mechanics with Hermitian Hamilton
operator, rigid phases of its eigenfunctions, and 
decay widths obtained from poles of the $S$ matrix.  
This can be seen also in the following manner. The time, the wave spends inside
the system at the energy of a resonance state, 
can be expressed by the Wigner time delay function which 
is proportional to  the width of the state.  
Numerical calculations performed for a quantum billiard with overlapping
resonance states by using the non-Hermitian Hamilton operator (\ref{heff})
show that the spectroscopic redistribution processes  
can be seen, indeed, also in the time delay function \cite{delay}.  
There is almost no time delay in the energy range
of a short-lived state while it is large at the energies of the
trapped states.  In the standard quantum mechanics with Hermitian Hamilton
operator, the spectroscopic redistribution processes are not involved.
Therefore, short-lived (aligned) resonance states do not appear,
the delay time can not be reduced and the transmission time 
can not be shortened.

\section{Conclusions}
 
The quantum brachistochrone problem of a physical system
can be studied by considering 
the time  needed for the transmission through the system  
from one of the attached leads to another one.
According to $S$ matrix theory, the transmission time at a certain energy 
$E$ is determined by the lifetime of the resonance states lying
at this energy. The lifetime of a resonance state is bounded from below:
it can not be smaller than allowing  traveling  
through the system in accordance with traveling through the attached leads,
i.e. the system may become  transparent at most \cite{comm2}
The difference between Hermitian and non-Hermitian quantum systems is that 
this lower bound can be reached in non-Hermitian systems by aligning the
wave functions of the system with those of the environment
while such a possibility does not exist in Hermitian systems.

The condition that an alignment of  wave functions of the system with those
of the environment becomes possible in non-Hermitian quantum mechanics,
is resonance overlapping such that many branch  points
exist in the parameter range considered.
Only in the neighborhood of these branch  points, 
the eigenfunctions $\phi_\lambda$
of the non-Hermitian Hamilton operator $H_{\rm eff}$
are really biorthogonal and have the possibility to align with the
traveling waves $\xi_C^E$ in the attached leads
due to their interaction via the continuum. Mathematically, the
alignment is a consequence of the normalization of the 
eigenfunctions of the non-Hermitian Hamiltonian $H_{\rm eff}$
according to (\ref{biorth1}).
The  alignment takes place in a hierarchical manner \cite{rep}.
It is maximal when many levels are almost aligned.
In this case, $\rho \approx 0$ in a certain range of the considered parameter
and the transmission is plateau-like with
$|t| \approx 1$ (for numerical examples see \cite{brsphas,brscorr}).
The system is  transparent, up to some dips  that appear
in the case when the system has many levels. These dips are
caused by the long-lived trapped resonance states that always
appear together with the short-lived aligned resonance states (due to widths
bifurcation) in the neighborhood of the branch points.
An example are the whispering gallery
modes in quantum billiards of Bunimovich type \cite{naz,brscorr}. 
At the critical point at which the
number of aligned states is exactly equal to the number of traveling waves
$\xi_C^E$ in the leads, $|\rho | > 0$ and $|t| <1$.   
 
This freedom to align the wave functions of the individual states
with the traveling waves $\xi^E_C$ in the attached leads does not exist 
in the Hermitian quantum mechanics.
Instead, the normalization of the wave functions
according to $\langle \phi_\lambda | \phi_\lambda \rangle =1$  fixes 
the phases of the individual wave functions in the 
Hermitian quantum mechanics  and prevents any alignment.
As a consequence, it is always $|r_\lambda | =1$ and 
$|\rho | = 1$ in standard  Hermitian quantum mechanics. 
The transmission takes place via waves standing at a certain energy   
for a certain time
inside the system. This time is longer than the traveling time, 
generally. The transmission  shows a characteristic resonance structure
that is described, in the standard theory, by means of the poles of the $S$
matrix. This resonance picture
can be seen in the regime of weak coupling between system and environment
where the individual resonances are
well isolated from one another as well as in the regime of strong coupling
where  narrow resonances are superposed by a smooth background. 
The crossover between these two borderline cases can {\it not} be described 
by standard Hermitian quantum mechanics as it is very well known in the
physics of open quantum systems. For example, an interpolation procedure
between these two limiting cases is proposed in \cite{shapiro}.
The reason for the failure of the Hermitian quantum mechanics in this case is,
as shown above, $|r_\lambda| <1$ and  $|\rho | <1$
in the crossover regime.

Summarizing it can be stated that the brachistochrone problem
is observable in realistic open quantum mechanical systems.  
It is directly related to the branch  points
of the non-Hermitian Hamilton operator. 
In the present paper, the transmission through a small open quantum 
billiard is considered as an example. The system becomes transparent in the 
regime of overlapping resonances
since the Hamilton operator $H_{\rm eff}$ is really non-Hermitian
in this regime and many branch points exist in the complex plane.

\vspace{.5cm}

{\bf
Acknowledgments} Valuable discussions with M. Berry,
E. Bulgakov, U. G\"unther, 
A. Sadreev and B. Samsonov are gratefully acknowledged.


\begin{thebibliography}{99}
\bibitem{comm2} In a 
many-particle system, the life time of a resonance state can not
be smaller than the lifetime of a single-particle resonance with the
corresponding quantum numbers.
\end{thebibliography}
\end{document}